\documentclass[10pt, a4paper]{article}

\usepackage{styles/arxiv}         
\usepackage[utf8]{inputenc}

\usepackage[english]{babel}
\usepackage[T1]{fontenc}    % use 8-bit T1 fonts
\usepackage{hyperref}       % hyperlinks
\usepackage{url}            % simple URL typesetting
\usepackage{booktabs}       % professional-quality tables
\usepackage{amsfonts}       % blackboard math symbols
\usepackage{microtype}      % microtypography
\usepackage{amsmath}
\usepackage{amssymb}        % maths symbol
\usepackage{amsthm}         % Proof environment
\usepackage{dsfont}         % For indicator function
\usepackage{float}
\usepackage{graphicx}
\usepackage{stmaryrd}       % for integer intervals llbracket
\usepackage{multirow}
\usepackage{tikz}

\usepackage{pgfplots}
\usepgfplotslibrary{colormaps}

\usepackage{csquotes}
\usepackage{todonotes}
\usepackage{soul}           % Strike through
\usepackage{marvosym}       % For smiley with "DejaVu Sans" font
\usepackage{caption}
\usepackage{subcaption}

\usepackage{algorithm}
\usepackage{algpseudocode}

\usetikzlibrary{trees}
\usetikzlibrary{arrows,automata}

\usepackage[edges]{forest}

\forestset{
  direction switch/.style={
    for tree={edge+=thick, font=\sffamily},
    where level>=1{folder, grow'=0}{for children=forked edge},
    where level=3{}{draw},
  },
}

\let\oldReturn\Return
\renewcommand{\Return}{\State\oldReturn}
\usepackage[citestyle=nature]{biblatex}
\pgfplotsset{compat=1.18}
\pgfplotsset{
    colormap={uni}{rgb255(0cm)=(0, 170, 220); rgb255(1cm)=(255, 20, 11)}
}

\usepgfplotslibrary{fillbetween}

\definecolor{BlueUnilu}{RGB}{0, 170, 220}
\definecolor{RedUnilu}{RGB}{255, 20, 11}

\graphicspath{ {./images/} }

\title{Statistically Enhanced Learning: a feature engineering framework to boost (any) learning algorithms}

\date{} 					% Or removing it

\author{
 Florian Felice \\
 Department of Mathematics\\
 University of Luxembourg\\
 \href{mailto:florian.felice@uni.lu}{\texttt{florian.felice@uni.lu}}\\
 \And
 Christophe Ley\\
 Department of Mathematics\\
 University of Luxembourg\\
 \href{mailto:christophe.ley@uni.lu}{\texttt{christophe.ley@uni.lu}}\\
 \And
 Stéphane Bordas\\
 Department of Engineering\\
 University of Luxembourg\\
 \href{mailto:stephane.bordas@uni.lu}{\texttt{stephane.bordas@uni.lu}}\\
 \And
 Andreas Groll\\
 Department of Statistics\\
 University of Dortmund\\
 \href{mailto:groll@statistik.tu-dortmund.de}{\texttt{groll@statistik.tu-dortmund.de}}\\
}

\begin{document}

\tikzstyle{every node}=[draw=black,thick,anchor=west]
\tikzstyle{selected}=[draw=red,fill=red!30]
\tikzstyle{optional}=[dashed,fill=gray!50]

\maketitle

\begin{abstract}
    Feature engineering is of critical importance in the field of Data Science.
    While any data scientist knows the importance of rigorously preparing data to obtain good performing models, only scarce literature formalizes its benefits.
    In this work, we will present the method of Statistically Enhanced Learning (SEL), a formalization framework of existing feature engineering and extraction tasks in Machine Learning (ML).
    The difference compared to classical ML consists in the fact that certain predictors are not directly observed but obtained as statistical estimators.
    Our goal is to study SEL, aiming to establish a formalized framework and illustrate its improved performance by means of simulations as well as applications on real life use cases.
\end{abstract}

\paragraph{Significance statement}

Statistically Enhanced Learning (SEL) is a promising approach to improving learning performance.
This work provides a formal definition of SEL and presents a framework for understanding its different components.
The framework identifies the intersections between Statistics, Enhanced (data processing), and Learning, and defines different levels of SEL features.
This work will enable researchers and practitioners to better understand and apply SEL to a wide range of learning problems.

% Structure from: https://www.nature.com/articles/s41467-022-33957-8

\subsubsection*{Introduction}

In the field of Machine Learning (ML), the preparation and pre-processing of the data is often considered equally or even more important than the model itself.
Students in data science are usually taught that 80\% of the workload on an ML project is about preparing the data, while the remaining 20\%  are concerned with the actual  choice of ML model \parencite{press2016cleaning}.
% \parencite{zhang2003data}
This is in sharp contrast to the focus put on the modeling part in comparison to the data preparation and its benefit to models.
As an illustration, the top 15 questions on Stack Overflow for the keywords "Machine learning" count 57.5 times more views than the top questions for "Data preparation" or "Data engineering"\footnote{"Machine learning" counts 831,126 views covering 94 answers while ``Data preparation'' counts 14,591 views and 18 answers. Accessed on June 29\textsuperscript{th}, 2023.}.

In this work, we therefore introduce Statistically Enhanced Learning, abbreviated SEL, which is a statistical feature engineering framework that allows building new features/covariates which cannot be directly observed.
The idea  is to enhance the performance of existing learning algorithms by extracting specifically targeted information that is not directly given by the data.
This allows adding the information of an unobserved or mis-measured signal under the form of a statistical covariate with a clear meaning. As we shall demonstrate, SEL works for any type of data (tabular, computer vision, text) and is a general approach to improve any learning algorithm.
We refer to learning as the general term, since it considers the large spectrum of data driven learning techniques (from classical statistical to advanced deep learning models).
As we will see, contributions from different domains (statistics, machine learning, econometrics, computer science, \ldots) have already unknowingly used SEL as feature engineering technique.
By our formalizing framework we will thus reunite and structure seemingly distinct approaches, which will shed new light on feature engineering.

We distinguish three levels of SEL features with increasing technical complexity:
\begin{enumerate}
    \item[SEL 1 -] \underline{Proxies}: addition of one or several features to represent another variable we cannot observe or do not have available. In statistics and econometrics, a proxy is a variable which is correlated with and used in place of an omitted variable \parencite{wooldridge2015introductory}. It can be a weak representation of the original signal, but still carries enough information from the unmeasured variable \parencite{Montgomery_2000}. An illustrative example from econometrics would be the proxy feature household consumption for the abstract and not-measurable concept of standards of living.
    \item[SEL 2 -] \underline{Descriptive statistics}: transformation of some existing features with classical statistical tools (e.g., count, moments, quantiles, \ldots) to summarize information in a meaningful way. Such summaries are particularly relevant with a large amount of predictors who, on their own, would be meaningless to add. For example, in the prediction of sports matches between two teams of, say, 30 players each, the average age is a useful predictor contrary to individual ages of players.
    \item[SEL 3 -] \underline{Advanced modeling features}: higher level of abstraction to extract information from available variables (that cannot be used as predictors themselves) via more advanced statistical tools (e.g., maximum likelihood estimators, causal estimands, moving averages, \ldots). It is important for the resulting covariates to bear a statistical nature, meaning that their uncertainty should be quantifiable and they should have  a concrete meaning to users. These variables should also add new information to the model to enhance its learning. Hence, this excludes dimensionality reduction techniques such as principal component analysis.
    As an illustrative example, the forecasting of wind energy production can be improved by adding exponentially weighted moving averages (EWMA) \parencite{Holt04} of wind speed measurements over the past 7, 14, 21 and 28 days.%In our case this would represent the “Abilities” variable estimated assuming a Poisson distribution.
\end{enumerate}

SEL 3 may require some further specifications.
% Modeling real-life phenomena by means of flexible probability distributions, with parameters bearing clear roles, is becoming an ever more important task nowadays due to the occurrence of numerous complex data sets \parencite{Ley21}.
% After having modeled, for example, a set of available variables with such a probability distribution and estimated its parameters, SEL 3 consists in constructing as new predictor a meaningful combination of these estimated parameters.
% A concrete example shall be given below.
Modeling real-life phenomena by means of flexible probability distributions, is becoming an ever more important task nowadays due to the occurrence of numerous complex data sets \parencite{Ley21}.
The parameters of such a probability distribution bearing clear roles, SEL 3 consists in constructing as new predictor a meaningful combination of these estimated parameters.
A concrete example shall be given below.
In causal inference, an estimand is a functional of the data distribution, which defines the target quantity of a statistical analysis.
It defines the quantity one wishes to estimate to answer a scientific question.

The borders between the different levels of SEL features are not strict and can even sometimes be quite thin.
For instance, depending on the probability distribution chosen in SEL 3, quantities from SEL 2 could be the resulting maximum likelihood estimators (MLEs), consider for instance the sample average being the MLE for the location parameter of a Gaussian distribution.
In the same vein, one could argue that the above-mentioned average player age (SEL 2) can be seen as a proxy (SEL 1) for a team's maturity (which is, in fine, what one tries to embed with this covariate).

To further understand the nature of SEL and the differences between the three levels, let us now look at the difference between classical learning and SEL from the perspective of Granger causality:
\parencite{granger1969investigating} defines a causal relation when a predictor of a phenomenon contains information that cannot be retrieved from another predictor.
In other words, if all predictors on earth to predict a phenomenon $\mathbf{Y}$ are contained in our input matrix $\mathbf{X}$, then we can consider the relation to be causal.

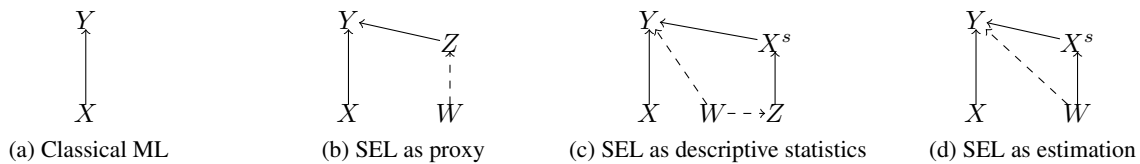
\begin{figure}[!ht]
     \centering
     \begin{subfigure}[a]{0.24\textwidth}
         \centering
         \begin{tikzpicture}[every node/.style={inner sep=0,outer sep=0}]
    \node[](Y) {$Y$};
    \node[below=1cm of Y](X) {$X$};
    \draw[->] (X) to (Y);
\end{tikzpicture}
         \caption{Classical ML}
         \label{fig:causal_classic}
     \end{subfigure}
     \hfill
     \begin{subfigure}[a]{0.24\textwidth}
         \centering
         \begin{tikzpicture}[every node/.style={inner sep=0,outer sep=0}]
    \node[](Y) {$Y$};
    \node[below=1cm of Y](X) {$X$};
    \node[right=1cm of X](W) {$W$};
    \node[above=0.7cm of W](Z) {$Z$};
    
    \draw[->] (X) to (Y);
    \draw[->] (Z) to (Y);
    \draw[->, dashed] (W) to (Z);
    % \draw[->, dashed] (W) to (Y);
\end{tikzpicture}
         \caption{SEL as proxy}
         \label{fig:causal_proxy}
     \end{subfigure}
     \hfill
     \begin{subfigure}[a]{0.24\textwidth}
         \centering
         \begin{tikzpicture}[every node/.style={inner sep=0,outer sep=0}]
    \node[](Y) {$Y$};
    \node[below=1cm of Y](X) {$X$};
    \node[right=0.5cm of X](W) {$W$};
    \node[right=0.5cm of W](Z) {$Z$};
    \node[above=0.7cm of Z](XS) {$X^s$};
    
    \draw[->] (X) to (Y);
    \draw[->] (XS) to (Y);
    \draw[->] (Z) to (XS);
    \draw[->, dashed] (W) to (Z);
    \draw[->, dashed] (W) to (Y);
    % \draw[->, dashed] (W) to (Y);
\end{tikzpicture}
         \caption{SEL as descriptive statistics}
         \label{fig:causal_descstat}
     \end{subfigure}
     \hfill
     \begin{subfigure}[a]{0.24\textwidth}
         \centering
         \begin{tikzpicture}[every node/.style={inner sep=0,outer sep=0}]
    \node[](Y) {$Y$};
    \node[below=1cm of Y](X) {$X$};
    \node[right=1cm of X](W) {$W$};
    \node[above=0.7cm of W](XS) {$X^s$};
    
    \draw[->] (X) to (Y);
    \draw[->] (XS) to (Y);
    \draw[->] (W) to (XS);
    \draw[->, dashed] (W) to (Y);
\end{tikzpicture}
         \caption{SEL as estimation}
         \label{fig:causal_estimate}
     \end{subfigure}
     \caption{Representation of SEL variables in a causal view vs. classical approaches.}
     \label{fig:causality_granger}
\end{figure}

%: ``\textit{If some other series $Y_t$, contains information in past terms that helps in the prediction of $X_t$, and if this information is contained in no other series used in the predictor, then $Y_t$ is said to cause $X_t$}''.
%\todo{I should rewrite this to paraphrase and cite the quote. Stéphane found that the series $Y_t$ comes a bit from nowhere}
% \todo[inline]{or \parencite{granger1980testing}: ``\textit{suppose that X and Y are the only two random variables in the universe and that a strong correlation is observed between them. [...] Suppose one tells that X does not cause Y leaving open the possibility of Y causing X. In the circumstances, the strong observed correlation might lead to acceptance of the proposition that Y does cause X.}''}
Putting this statement in the context of SEL, we can summarize the workflow with the diagram from Figure \ref{fig:causality_granger}.

The classical (modeling and) learning approaches as represented in Figure \ref{fig:causal_classic} consider an input matrix $\mathbf{X}$ that influences the target variable $\mathbf{Y}$.
The modeler then tries to estimate the relation $\mathbf{Y} = f(\mathbf{X})$, with $f$ being any (potentially nonlinear) transformation of the input.
With Granger's logic \parencite{granger1969investigating}, if $\mathbf{X}$ contains all variables  influencing $\mathbf{Y}$ and we know that $\mathbf{Y}$ does not cause $\mathbf{X}$, then the relation between $\mathbf{X}$ and $\mathbf{Y}$ can even be considered as causal.
SEL comes at play in the opposite situation, when $\mathbf{X}$ does not contain all the signals influencing the target $\mathbf{Y}$ (see Figures \ref{fig:causal_proxy}-\ref{fig:causal_estimate}).
Instead, we know that other factors $\mathbf{W}$ have a direct influence on $\mathbf{Y}$ but they cannot be observed.
The modeler then uses new substitute variables, denoted $\mathbf{X}^s$ in the diagram, to represent the missing signal.
In other words, since the relation $\mathbf{Y} = f(\mathbf{X}, \mathbf{W})$ cannot be explicitly written because $\mathbf{W}$ is not observed, the modeler substitutes $\mathbf{W}$ by $\mathbf{X}^s$ and focuses on estimating the relation $\mathbf{Y} = f(\mathbf{X}, \mathbf{X}^s)$.
The three levels of SEL correspond to the following representations.

\begin{enumerate}
    \item The dashed link in Figure \ref{fig:causal_proxy} exists but cannot be observed so the scientist has to use $\mathbf{Z}$ as an alternative source of information to model the phenomenon $Y$.
    In situations such as household consumption used to represent unobserved standard of living \parencite{Montgomery_2000}, the substitute variables are proxy variables. \cite{wooldridge2009estimating} enforces this practice by underlining the benefit of such proxies.
    \item The link between the variable of interest $\mathbf{W}$ and the target $\mathbf{Y}$ as illustrated in Figure~\ref{fig:causal_descstat} is indirect.
    % The variable $W$ is unobserved and replaced by the proxy variable $Z$.
    % The dashed line represents the link between the variables: a correlation.
    % The variable $Z$ is then directly included in the model where the plain line represents the adjunction to the model.
    In situations like sports predictions, we know that the information contained in $\mathbf{X}$ is not sufficient to model $\mathbf{Y}$ accurately.
    The maturity of players as the true missing signal $\mathbf{W}$ is too important to be ignored as a predictor. The players' ages (variable $\mathbf{Z}$) are a good indicator (proxy) for their maturity. However,  the ages of individual players cannot be used alone as predictors and would be meaningless to the model. Hence, instead the players' average age ($\mathbf{X}^s$) should be used.
    Therefore, SEL in $\mathbf{X}^s$ acts as a representative summary of the information contained in $\mathbf{Z}$.
    \item In the last situation, depicted in Figure \ref{fig:causal_estimate}, the signal $\mathbf{W}$ causing $\mathbf{Y}$ cannot be observed either but SEL is used to estimate the relation via $\mathbf{X}^s$.
    SEL is no longer a proxy with the goal to add information but instead an actual estimation of the missing signal $\mathbf{W}$.
    In a wind energy prediction example, we wish to input an estimate of the to-be-expected wind speed ($\mathbf{W}$ in Figure~\ref{fig:causal_estimate}) based on recent wind speed measurements.
    After defining the time window, EWMA allows obtaining such an estimate ($\mathbf{X}^s$) by weighting the recency of the individual measurements.
    The true to-be-expected wind speed of course cannot be observed.
\end{enumerate}

Based on this Granger causality discussion, we wish to stress again that the principle of SEL is not to add any information in order to marginally increase predictive performance as it is the case in linear regression with the coefficient of determination $R^2$ that mechanically increases even for variables that are just weakly related to $Y$.
SEL rather is a means to recover information from signals that cannot be detected.

%In other instances, extra information such as moments of color histogram in \cite{xuan_steganalysis_2005} or count of characters in \cite{senthil2018detecting} are added to the model via the SEL variables $X^s$.
% This is the matrix of SEL variables.
%Its relation with $Y$ becomes indirect since the information is (partially) captured via the proxy variables $X^s$.

%It is worth highlighting here that the relation between $X^s$ and $Y$ cannot be causal since we know that the true relation is drawn from $W$.
%Therefore $X^s$ only serves as (imperfect) mirror for information.
%This can be linked to the \textit{causal cross spectra} concept from \cite{granger1969investigating}.

\begin{figure}[!ht]
    \centering
    \begin{tikzpicture}[thick, set/.style = {circle, minimum size = 3cm}]
    % Sets
    \node[set,draw=none] (DataProcessing) at (0,0) {};
    \node[set,draw=none] (Statistics) at (1.8,0) {};
    \node[set,draw=none] (Learning) at (0.9,1.8) {};

    % Intersection
    \begin{scope}
        \clip (0,0) circle(1.5cm);
        \clip (1.8,0) circle(1.5cm);
        \clip (0.9,1.8) circle(1.5cm);
        \fill[RedUnilu](0,0) circle(1.5cm);
    \end{scope}

    % Circles outline
    \draw (0,0) circle(1.5cm);
    \draw (1.8,0) circle(1.5cm);
    \draw (0.9,1.8) circle(1.5cm);

    % Set intersection labels
    \node[align=left,draw=none] at (-1.5,0) {Data\\processing};
    \node[align=center,draw=none] at (0.2,2.2) {Learning};
    \node[align=right,draw=none] at (1.8,0) {Statistics};
    \node[align=center,draw=none] at (0.5,0.7) {SEL};

    % Arrows for intersections
    \draw[gray, <-,thick] (1.7,1) -- (3,1.8) node [right,draw=none] {Statistical learning};
    \draw[gray, <-,thick] (0.1,1) -- (-1.1,1.8) node [left,draw=none] {Feature engineering};
    \draw[gray, <-,thick] (0.9,-0.5) -- (0.9,-1.8) node [below,draw=none] {Data mining};
\end{tikzpicture}
    \caption{Statistically Enhanced Learning as the intersection between three fields} \label{fig:sel_fields}
\end{figure}
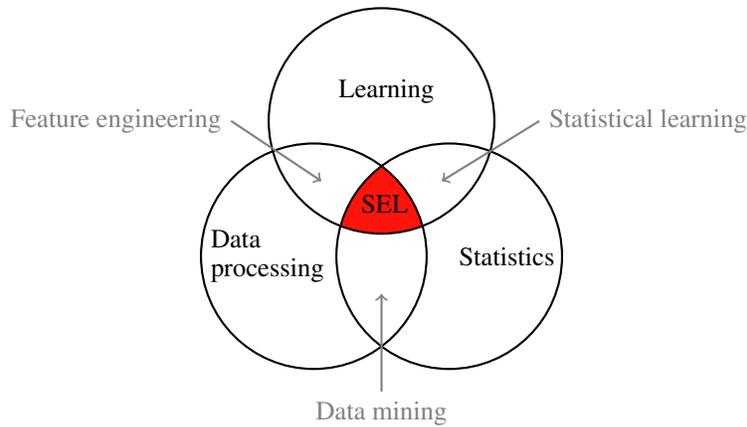

We shall now properly contextualize SEL within the domains of learning and data science, hereby contributing a new overview of various domains that is insightful \emph{per se}.
We define Statistically Enhanced Learning as inherited from three different fields:
\begin{itemize}
\item[$\bullet$] Learning: General field aiming to process some input and generate predictions. It can be as general as machine and deep learning as illustrated in Figure~\ref{learningVenn} below, which is taken from \cite{goodfellow2016deep}. Quoting \cite{hastie2009elements}: ``\emph{Using this data we build a prediction model, or learner, which will enable us to predict the outcome for new unseen objects. A good learner is one that accurately predicts such an outcome.''}
\item[$\bullet$] Enhanced (data processing): Field that includes data preparation steps to later enhance the learning performance. It can be  compared to information processing as defined by \cite{ralston2003data} and includes steps such as data cleaning and data preparation.
\item[$\bullet$] Statistics: In a wide sense, field that includes descriptive statistics, inference, modeling and causality.
\end{itemize}
We  visually represent these fields in Figure~\ref{fig:sel_fields}. As they are overlapping domains, we can identify their intersections as follows:
\begin{itemize}
\item[$\bullet$] Statistical learning = Learning $\cap$ Statistics: any learning algorithm that we work with (e.g., linear regression, Random forests, Neural networks). It is an interdisciplinary field by nature as it intersects with artificial intelligence and areas of engineering or other disciplines \parencite{hastie2009elements}.
\item[$\bullet$] Feature engineering = Learning $\cap$ Data processing: First and crucial part before the modeling exercise. It consists in the preparation of the data set by processing the input data (cleaning, scaling, etc.) \parencite{zheng2018feature}
\item[$\bullet$] Data mining = Statistics $\cap$ Data processing: Part of the literature which consists in extracting information/knowledge from data \parencite{chakrabarti2006data}. Extracted information can be used in inferential statistics, learning  or generated business knowledge and metrics.
\end{itemize}

\begin{figure}[!ht]
    \centering
       
\begin{tikzpicture}
        % left hand
        % \scope
        % \clip 
        %       (1,0) circle (1);
        % \fill (0,0) circle (1);
        % \endscope
        % % right hand
        % \scope
        % \clip 
        %       (0,0) circle (1);
        % \fill (1,0) circle (1);
        % \endscope
        % outline
        \draw (0,3) circle (3) (0,6)  node [text=black,below, draw=none] {AI}
              (0,2.5) circle (2.5) (0,5)  node [text=black,below, draw=none] {Learning}
              (0,2) circle (2) (0.5,4)  node [text=black,text width=2cm,below, draw=none] {Machine Learning}
              % (0,2) ellipse (1cm and 2cm) %[text=black,below] {Statistical Learning}
              (0,1) circle (1) (0.4,2)  node [text=black,text width=1.6cm,below, draw=none] {Deep Learning};
    
        \draw [-stealth] (0.2,5.7) -- (3.5,5.7);
        \draw [-stealth] (0.7,4.7) -- (3.5,4.7);
        \draw [-stealth] (1.2,4.1) -- (3.5,3.7);
        \draw [-stealth] (0.8,3.5) -- (3.5,2.5);
        \draw [-stealth] (0.8,1.5) -- (3.5,1.5);

        \draw[black, ->,thick] (0.2,5.7) -- (3.5,5.7) node [right,draw=none] {\footnotesize{Scripts, rule based systems, etc.}};
        \draw[black, ->,thick] (0.7,4.7) -- (3.5,4.7) node [right,draw=none] {\footnotesize{Algorithms learning from data}};
        \draw[black, ->,thick] (1.2,4.1) -- (3.5,3.7) node [right,draw=none] {\footnotesize{Linear modeling (OLS, logit, etc.)}};
        \draw[black, ->,thick] (0.8,3.5) -- (3.5,2.5) node [right,draw=none] {\footnotesize{Nonlinear modeling (Random Forests, XGBoost, etc.)}};
        \draw[black, ->,thick] (0.8,1.5) -- (3.5,1.5) node [right,draw=none] {\footnotesize{Neural Networks}};
        
        % \node [text width=3cm, draw=none] at (6.1,5.7) {\footnotesize{Scripts, rule based systems, etc.}};
        %\node [text width=5cm, draw=none] at (6.1,4.7) {\footnotesize{Algorithms learning from data}};
        %\node [text width=5cm, draw=none] at (6.1,3.7) {\footnotesize{Linear modeling (OLS, logit, etc.)}};
        %\node [text width=5cm, draw=none] at (6.1,2.5) {\footnotesize{Nonlinear modeling (Random Forests, XGBoost, etc.)}};
        %\node [text width=5cm, draw=none] at (6.1,1.5) {\footnotesize{Neural Networks}};
              % (-2,-2) rectangle (3,2) node [text=black,above] {$H$};
    \end{tikzpicture}
    \caption{Venn diagram for Artificial Intelligence and Learning (inspired from \cite{goodfellow2016deep})} \label{learningVenn}
\end{figure}
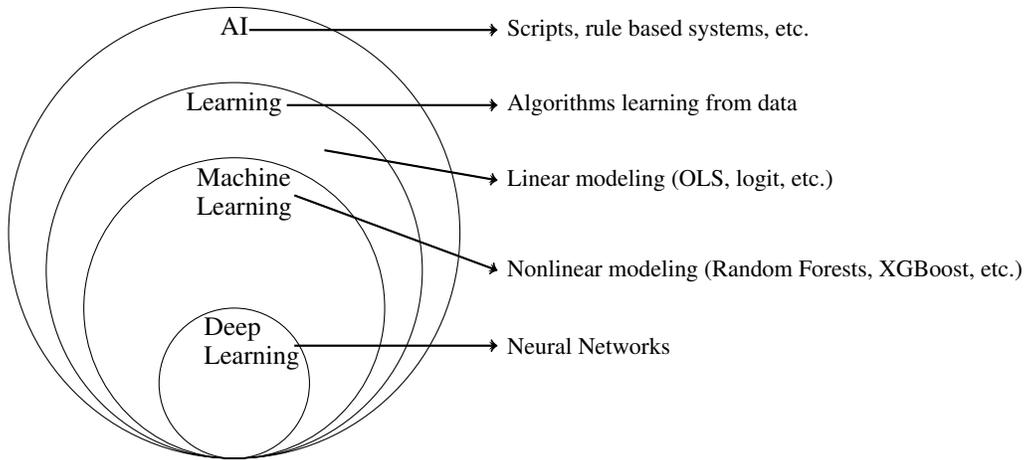

SEL is at the intersection of these fields as illustrated in Figure~\ref{fig:sel_fields}.
It consists of  augmenting a data set with easy-to-understand features from either SEL level in order to improve the performance of any learning algorithm.
It thus allows retrieving information about a missing signal. As they are not measured directly, SEL variables bear an extra layer of uncertainty, which one can quantify thanks to the requirement of SEL covariates to be of statistical nature.
We also note that SEL is  creating a bridge between the fields of statistics and machine learning which all too often are considered as competitors/distinct \parencite{breiman2001statistical}.

We shall now provide examples of contributions whose methodology falls under the umbrella of SEL.
We analyze these works under the formal SEL framework and, hence, shew new light on them.
To show the universal applicability of SEL, we will consider a subdivision based on the type of data.

\paragraph{Tabular linear setting}
%In a tabular setting, e
Econometricians often face the problem of unobserved variables to model complex phenomena.
Important variables to a model (such as income, willingness to pay, feelings) are either not available to the modeler or not measurable.
The use of proxy variables, hence SEL 1, is a solution to compensate the lack of necessary signals. As already mentioned, \cite{Montgomery_2000} analyze the benefit of the use of proxies such as household consumption from demographic surveys to represent standards of living and show that even weak proxy variables can still capture the desired signal from the unobserved feature.
They also claim that adding proxies can help reduce the inconsistency of the estimated parameters.

The two-stage least squares approach (aka. 2SLS, \cite{wooldridge2015introductory}) is another indirect modeling strategy which is used to incorporate a feature that cannot directly be included in the regression model when it contradicts the OLS assumptions (when the covariates $X$ are not independent from the residuals $\varepsilon$).
Therefore, including an instrumental variable (IV) as a two-stage approach is preferred by first regressing the conflicting variable $Z$ as a function of some regressors $W$, thus we regress $Z = \beta_W W + \varepsilon'$ for some error term $\varepsilon'$.
The estimated variable $Z$ then feeds the main regression model by $Y = \beta_X X + \beta_Z \hat{Z} + \varepsilon$, and this is clearly an instance of SEL 3.
% The first estimate $\hat{Z}$ is a statistical quantity, hence the overall approach can be referred to as SEL.
A generalization of IV methods \parencite{Chesher_2020} formalizes the framework which can be applied to discrete modeling or in the context of high heteroscedasticity.

It  also happens frequently that observed variables are noisy and therefore cannot be used as predictors.
In such cases, modelers pre-process those variables via statistical methods such as kernel composition/decomposition or Fourier transforms \parencite{Schennach_2020}.
Such techniques help deal with mis-measured variables (or measured with noise) and prepare a cleansed signal that will enhance the learning step.
This approach falls under the umbrella of SEL 3.

\paragraph{Tabular nonlinear setting}
\cite{lin_evaluation_2019} use moments of some longitudinal features to classify abnormal bitcoin network addresses.
They add the first four moments (namely mean, variance, skewness and kurtosis) from time-dependent variables to extract intrinsic information of the variable to classify bitcoin addresses. Their method outperforms existing models and shows the high importance of the moment-based variables, hence of SEL 2.

%Some can consist of predicting the future values of a series \cite{briand2022monitoring} to reflect the time dependency when monitoring athletes' injuries.
%The predicted values from the moving average are added as new features together with rolling means and other moments.

In the context of football goals prediction of national teams, \cite{groll2019hybrid, GrollHybridRF_2021} use a so-called \emph{hybrid} approach to estimate unobserved variables and augment the data set for a Random Forest model.
On the one hand, they build a novel covariate as the average age of players, which is of course SEL 2.
On the other hand, they add a statistical feature which aims to represent the strength of the two opponents.
To this end, they consider historical games of all national teams over an 8-year-period preceding the tournament whose matches they intend to predict and model the joint distribution of goals scored by home and away teams ($i$ and $j$, respectively) by the bivariate Poisson distribution.
Hereby, the parameters $\lambda_i$ and $\lambda_j$ represent the mean parameters of the Poisson process and are assumed to be of the form
$\log(\lambda_i) = \beta_0 + (r_i - r_j) + h \cdot \mathds{1}(\text{team } i \text{ playing at home})$,
where $\beta_0 \in \mathbb{R}$ is a common intercept
and $h \in \mathbb{R}$ is the effect of playing at home.
The real-valued parameters $r_i$ and $r_j$ are the strength parameters of the home team $i$ and away team $j$, and they are estimated by means of weighted maximum likelihood.
The weights are chosen such that more importance is given to more recent matches.
These estimated strength parameters are then included as a new covariate to the final model for predicting scores.
As shown in \cite{groll2019hybrid, GrollHybridRF_2021}, this approach helps reduce the RMSE and allows even outperforming the bookmakers, which are a golden standard in sports prediction.
This hybrid approach clearly falls in the category of SEL 3, and the authors also showed that the SEL 3 variables have the highest variable importance in their Random Forest model.

\paragraph{Computer vision}
To analyze and classify images, \cite{xuan_steganalysis_2005} build a data set only composed of moment features to determine whether an image contains a hidden message in a picture or not.
The moments are derived from the wavelet subbands of an image to represent the color histogram of a picture.
These extracted color distributions help create features (by means of moments) that are important covariates sensitive to the change of colors and help improve the detection of hidden messages.
This approach in computer vision falls under the umbrella of SEL 2.
Other contributions also extract moments from images \parencite{soranamageswari_statistical_2010} or temporal signals \parencite{borith_prediction_2020} for classification purposes.

\paragraph{Natural Language Processing}
In the field of text classification and analysis, the main challenge consists in capturing  information carried by the word tokens. Some techniques consist in counting characters in a text to create new features for the data set \parencite{senthil2018detecting, sogaard2015inverted}, which is part of SEL 2.
A more advanced approach of counting words, which however still falls under SEL 2, is the Term Frequency-Inverse Document Frequency (TF-IDF) technique \parencite{ramos2003using}, which weights the counts by the frequency of the words appearing in a corpus, thus representing the importance of a word.
Another popular approach to deal with textual inputs is Word2Vec \parencite{mikolov2013efficient}.
This semantic-based technique uses neural networks with embeddings to produce numeric representation of words in a high-dimensional vector space.
The trained model helps compare words with the vector of semantically similar words.
At first sight, Word2Vec does not appear to be part of SEL 3 as it is a complex machine learning feature extraction method, however very recently \cite{DSWS23} showed that, under a copula-based statistical model for text data, Word2Vec can be interpreted as a statistical estimation method for estimating the point-wise mutual information, hence qualifying it as part of SEL 3.

\cite{lilleberg2015support} uses a combination of TF-IDF and Word2Vec to classify text into defined sentiment categories.
In our framework, their approach can be perceived as a double enhancement, as an SEL 2 technique is applied on SEL 3 type features.

\begin{figure}[!ht]
    \centering
    % Tikz source https://texample.net/tikz/examples/filesystem-tree/

\begin{forest}
      % forest preamble: determine layout and format of tree
      direction switch
      [\textbf{Statistically Enhanced Learning}
        [\textbf{\textit{Images}}
          [\textit{Color histogram \parencite*{mutlag2020feature}}]
          [\textit{Wavelets \parencite*{xuan_steganalysis_2005}}]
          [\textit{Moments \parencite*{soranamageswari_statistical_2010}}]
          [\textit{Fourier transform \parencite*{li2002palmprint}}]
        ]
        [\textbf{\textit{Text}}
          [\textit{Bag of words \parencite*{tsai2012bag}}]
          [\textit{$n$-gram \parencite*{ahuja2019impact}}]
          [\textit{Character count \parencite*{senthil2018detecting}}]
          [\textit{\textbf{Embeddings} \parencite*{selva2021review}}
            [\textit{word2vec \parencite*{lilleberg2015support}}]
            [\textit{TF-IDF \parencite*{aizawa2003information}}]
          ]
        ]
        [\textbf{\textit{Tabular}}
          [\textbf{\textit{Mismeasured variables}}
            [\textit{Fourier transform \parencite*{Schennach_2020}}]
            [\textit{Kernel decomposition \parencite*{Schennach_2020}}]
            [\textit{GIV / 2SLS \parencite*{Chesher_2020, wooldridge2015introductory}}]
            % [\textit{Probabilistic Random Forest \parencite*{reis2018probabilistic}}]
          ]
          [\textbf{\textit{Unobserved variables}}
            [\textit{Proxy \parencite*{Montgomery_2000, wooldridge2009estimating}}]
            [\textit{MLE \parencite*{GrollHybridRF_2021, groll2019hybrid}}]
          ]
          [\textbf{\textit{Sequential}}
            [\textit{Moments \parencite*{lin_evaluation_2019}}]
            [\textit{Cumulants \parencite*{abu-romoh_automatic_2018}}]
            [\textit{Time Series model \parencite*{briand2022monitoring}}]
          ]
        ]
      ]
\end{forest}
    \caption{A list of Statistically Enhanced Learning methods by type of data} \label{fig:formalize_tree}
\end{figure}
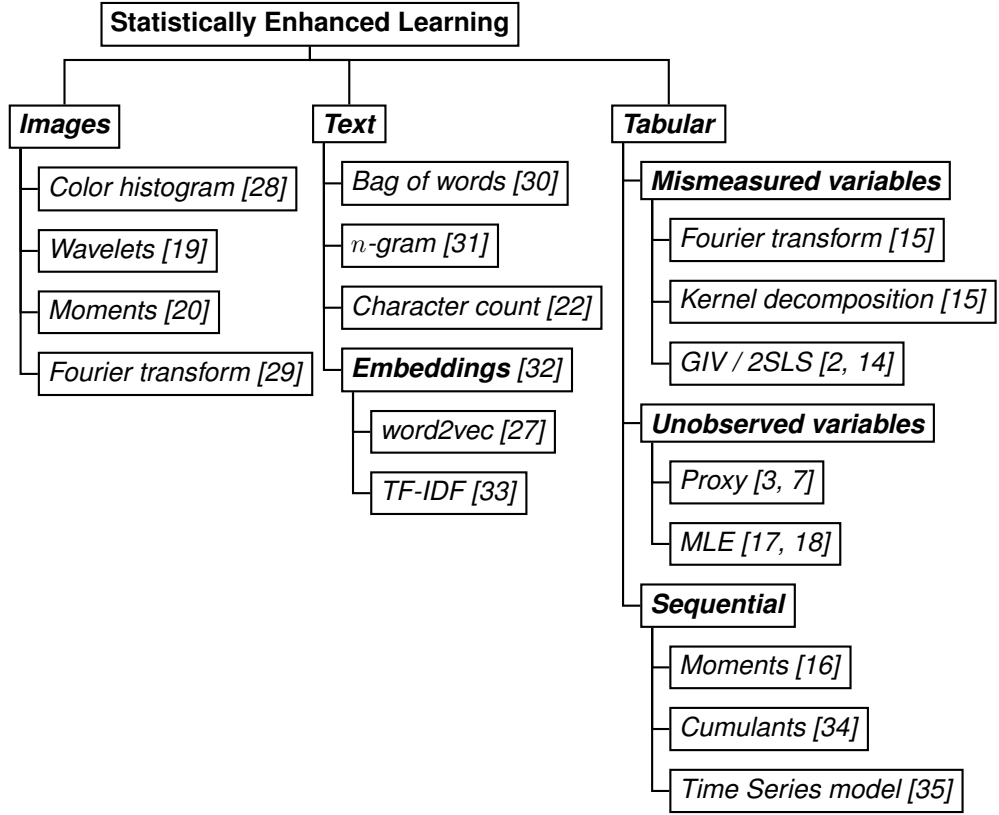

\paragraph{A unifying framework} As these examples show, our proposed Statistically Enhanced Learning is a general framework that gives a structure to hitherto distinct approaches.
For illustrative purpose, we summarize them in the diagram from Figure \ref{fig:formalize_tree}. 

Somewhat related to SEL is the recently proposed Probabilistic Random Forest \parencite{reis2018probabilistic}, which sets itself in the field of mismeasured variables. It is an adaptation of Breiman's Random Forest \parencite{breiman2001random} to account for the noise of measured features.
It considers quadruplets of the form $(x_i, \Delta x_i, y_i, \Delta y_i)$ instead of the usual pair $(x_i, y_i)$, where $\Delta x_i$ (resp., $\Delta y_i$) represents the uncertainty when measuring $x_i$ (resp. $y_i$).
In particular, the authors assume that each observed value is drawn from some normal distribution where $X_i \sim \mathcal{N}(x_i, \Delta x_i)$, so the additional quantity $\Delta x_i$ is added to the model and can be considered as a statistically estimated quantity.
Indeed, in fields such as astronomy, data often come from multiple sources (e.g., satellites) where the same observation is measured by different instruments.
The measure then contains uncertainty.
Not only did the authors include this additional source of information, but they adapted the Random Forest logic to account for this uncertainty.
The split from a node in a tree depends on this quantity $\Delta x_i$ and is no longer a boolean true or false.
This gives the model probabilistic considerations that improve its performance when uncertainty in measurements increases, but also allows deriving probability distributions of the target.

% ## APPLICATIONS
%\section{Illustrative applications}\label{sec:applications}

So far we have defined Statistically Enhanced Learning, presented its detailed structure, contextualized it within the realm of Data Science and Artificial Intelligence, and showed how existing approaches from the literature are embraced by SEL.
Next, we will demonstrate the learning performance enhancement of SEL by means of various examples, starting with synthetic data.

\subsubsection*{Benchmarking with simulated data}\label{sec:simulations}

%\todo[inline]{Re-write paragraph with latest results. Add SHAP results. Add results with limitation example?}

By means of Monte Carlo simulations, we will compare the performance of ML models with SEL covariates versus regular ML models. For our simulations, we consider $n = 1,500$ observations and $p$ predictors, whose values are simulated from a Gaussian distribution.  
Before computing the response variable, we generate some underlying process $\mathbf{Z}_{i}, i=1,\ldots,n$ of length $m=400$ for each of the $n$ individuals.
This process follows a Cauchy distribution whose parameter $\mu$ (the location parameter) will directly constitute a variable in the data set.
Formally, for an individual $i$, the regression function writes
\begin{equation}\label{eq:simu_reg_model}
    \mathbf{Y}_i = \beta' \mathbf{X}_i + \beta_\mu \mathbf{\mu}_i^2 + \mathbf{\varepsilon}_i
\end{equation}
where $\mathbf{X}_i \in \mathbb{R}^p$ is the $p$-dimensional vector of observed covariates, $\varepsilon_i \sim \mathcal{N}(0, 1)$ is the residual term, $\mathbf{\mu}_i$ is the location parameter of the underlying Cauchy process $\mathbf{Z}_i$ and $\beta \in \mathbb{R}^p$ and $\beta_\mu \in \mathbb{R}$ are the parameters to be learnt/estimated. We  assume that we cannot observe the parameter $\mu_i$ but only the underlying process $\mathbf{Z}_i$.

To learn the regression function (\ref{eq:simu_reg_model}) from data, we prepare three models using XGBoost \parencite{chen2016xgboost}.
A first baseline model considers that we can only observe the matrix of $p$ covariates $\mathbf{X}$.
A second model -- that we denote ``SEL 2'' -- uses the underlying process $\mathbf{Z}$ and computes the empirical mean, so that for each individual $i$ we have $X^s_i = m^{-1} \sum_{j=1}^{k} Z_{i,j}$.
A last model -- denoted ``SEL 3'' -- estimates the parameter of the Cauchy distribution via MLE from the underlying process $\mathbf{Z}$. We then have $X^{s}_{i} = \hat{\mu}_i$.
Since we consider that we do not know the form of the actual variable in the model, we input the estimated parameter $X^{s}_{i}$ with no further transformation as an additional variable into the XGBoost algorithm.

For replicability purposes, the full details on the simulations and data generation can be found in the repository referred in the \hyperref[sec:code]{Code availability Section}.

%For our simulations, we generate $p$ random variables from some probability distributions (Poisson, Gaussian, etc.) for a fixed number of individuals $i = 100$.
%We concatenate the $p$ variables from all of the $i$ individuals to construct one data set.
%We compute the response variable $Y$ to be predicted as a complex nonlinear combination of the $p$ generated variables.
%To train our model, we will omit $m$ covariates that we consider as unobserved.
%Here, we have $m = 1$.

%To cope for the missing information of $m$ unobserved variables, we estimate $s$ new variables as our SEL features.
%We compute the mean of the variable $Y$ for each individual which goal is to encode the information of the individual.

%Finally, we train a first ML model (here Random Forests \cite{breiman2001random}) on $p - m$ variables as a regular (vanilla) model.
%A second Random Forest model is trained with the augmented dataset composed of $p - m + s$ covariates.

\begin{figure}[!h]
    \centering
    \begin{tikzpicture}
    \begin{axis}[
    line width=0.5,
    x tick label style={font=\normalsize, draw=none, anchor=north},
    y tick label style={font=\normalsize, draw=none, anchor=east},
    xlabel={Number of columns $p$},
    ylabel={Relative RMSE versus baseline (base 100)},
    label style={font=\normalsize, draw=none},
    legend style={at={(0.95, 0.3)}, anchor=north east, fill=none},
    %y tick label style={
    %    /pgf/number format/.cd,
    %    fixed,
    %    precision=2
    %},
    ]
        % Performance for SEL
        \addplot[teal] table [x=n_cols, y=sel_mean, col sep=comma] {images/data/SimuResultsCauchy.csv};
        \addlegendentry[draw=none]{SEL (MLE)};
        % Confidence interval for SEL
        \addplot+[name path=SEL_p5, lime!50, mark=none, forget plot] table [x=n_cols, y=sel_p5, col sep=comma] {images/data/SimuResultsCauchy.csv};
        \addplot+[name path=SEL_p95, lime!50, mark=none, forget plot] table [x=n_cols, y=sel_p95, col sep=comma] {images/data/SimuResultsCauchy.csv};
        \addplot+[lime!50, forget plot] fill between[of=SEL_p5 and SEL_p95];
    
        % Performance for Moments
        \addplot[red] table [x=n_cols, y=moments_mean, col sep=comma] {images/data/SimuResultsCauchy.csv};
        \addlegendentry[draw=none]{SEL (moments)}
        % Confidence interval for Moments
        \addplot+[name path=mmt_p5, red!25, mark=none, forget plot] table [x=n_cols, y=moments_p5, col sep=comma] {images/data/SimuResultsCauchy.csv};
        \addplot+[name path=mmt_p95, red!25, mark=none, forget plot] table [x=n_cols, y=moments_p95, col sep=comma] {images/data/SimuResultsCauchy.csv};
        \addplot+[red!25, forget plot] fill between[of=mmt_p5 and mmt_p95];
        % Plot baseline
        \addplot[blue] table [x=n_cols, y=vanilla, col sep=comma] {images/data/SimuResultsCauchy.csv};
        
        \addlegendentry[draw=none]{Baseline}
    \end{axis}
\end{tikzpicture}
    \caption{Relative RMSE of SEL with $s = 1$ generated feature versus baseline approach (the lower the better)}
    \label{fig:simu_perf}
\end{figure}
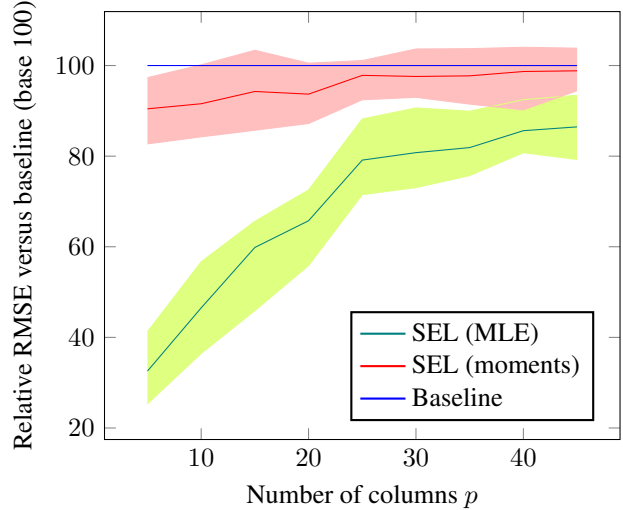

We report in Figure \ref{fig:simu_perf} the ratio of RMSE for the baseline model versus the SEL approaches as a function of the number of variables $p$.
Simulations are run $10,000$ times to derive values and credible intervals.

\begin{figure}[!ht]
     \centering
     \begin{subfigure}[a]{0.45\textwidth}
         \centering
         \pgfplotstableread[col sep = comma]{images/data/SimulationsFeatImport.csv}\tabledm
\def\xlistmacro{}
\def\xliststring{}
\pgfplotstableforeachcolumnelement{feature}\of\tabledm\as\entry{%
\xifinlist{\entry}{\xlistmacro}{}{
        \listxadd{\xlistmacro}{\entry}
        \edef\xliststring{\xliststring\entry,}
    }
}
\begin{tikzpicture}
    %\begin{axis}[
    % xlabel={Label},
    %ylabel={Count},
    %symbolic x coords/.expand once={\xliststring},
    %xmajorgrids=true,
    %x tick label style={rotate=60,anchor=east}
        %]
        %\addplot [ybar] table [x=feature, y=importance, col sep=comma] {\tabledm};
        %\end{axis}
        \begin{axis}[
            bar width=5pt,
            x tick label style={font=\footnotesize, draw=none, anchor=north},
            y tick label style={font=\footnotesize, draw=none, anchor=east},
            label style={font=\footnotesize, draw=none},
            width=\linewidth, % changed size a bit
            height=0.85\linewidth,
            xbar, % not xbar interval
            %bar width=12pt,
            % xmin=0.1, % does it make more sense to start at zero?
            % xmax=1.1, % reduced this a bit
            xlabel = Importance,
            axis x line=bottom,
            axis y line=left,
            enlarge y limits=0.01, % no need to enlarge x limits
            symbolic y coords/.expand once={\xliststring},
            ytick = data,
        %    y tick label style={rotate=45, anchor=east}, % I wouldn't use this
            %xticklabel style={
            %  /pgf/number format/fixed % no scientific notation for the smallest values
            %}
         ]
            \addplot[BlueUnilu, fill=BlueUnilu] table [y=feature, x=XGB_SEL, col sep=comma] {\tabledm};
        \end{axis}
\end{tikzpicture}
         \caption{Feature importance}
         \label{fig:simu_feat_imp}
     \end{subfigure}
     \hfill
     \begin{subfigure}[a]{0.45\textwidth}
         \centering
         \begin{tikzpicture}
    \begin{axis}[
    scatter/classes={a={mark=o,draw=red}},
    width=\linewidth,
    x tick label style={font=\footnotesize, draw=none, anchor=north},
    y tick label style={font=\footnotesize, draw=none, anchor=east},
    xlabel={SEL variable value},
    ylabel={SHAP value},
    label style={font=\footnotesize, draw=none},
    legend style={at={(0.95, 0.3)}, anchor=north east, fill=none},
    %y tick label style={
    %    /pgf/number format/.cd,
    %    fixed,
    %    precision=2
    %},
    ]
        % Partial Dependence plot
        \addplot[scatter,only marks, red] table [x=mu, y=shap_mu, col sep=comma] {images/data/SimuResultsPDP.csv};
        
    \end{axis}
\end{tikzpicture}
         \caption{Partial Dependence Plot}
         \label{fig:simu_pdp}
     \end{subfigure}
     \caption{Analysis of the model with TreeSHAP for the SEL model with $p=10$ variables}
     \label{fig:simu_shap_analysis}
\end{figure}
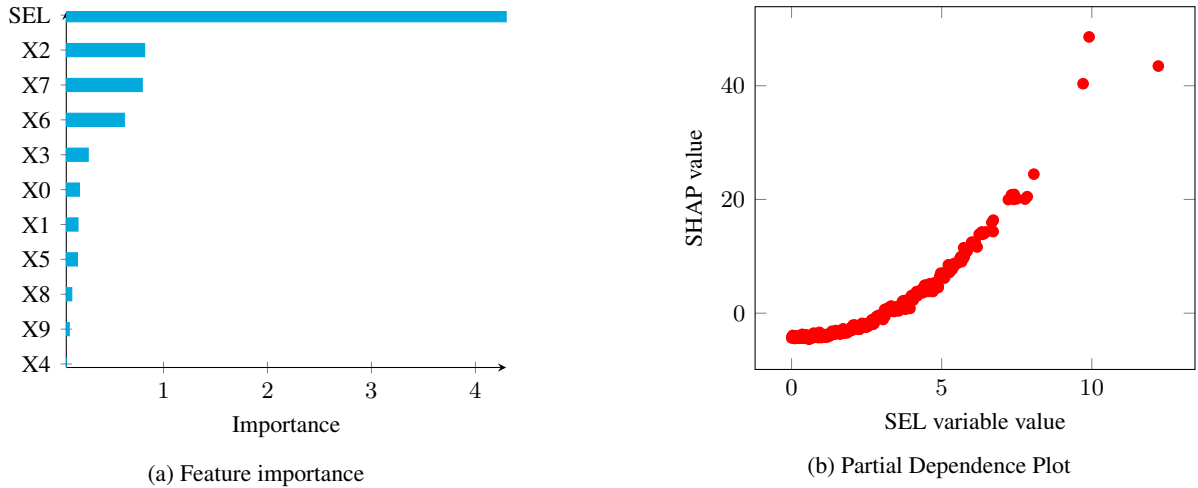

We can observe that the performance of the SEL models tends to be consistently better than the baseline approach, in particular when only few observable variables are present in the model. This suggests that the relative importance of the SEL features is quite high compared to the other covariates. We also analyzed the performance of the XGBoost model with the SEL 3 variable for one iteration with $p=10$. The formula used to generate our data in this case  is defined by
\begin{equation}\label{eq:simu_formula_example}
    Y = -1.04 X_0 -1.32 X_1 + 4.50 X_2 -1.69 X_3 + 0.53 X_4 + 1.34 X_5 + 3.35 X_6 + 4.10 X_7 -0.99 X_8 + 0.98 X_9 + 4.50 \mu^2 + \varepsilon.
\end{equation}
When analyzing the feature importance of the model using TreeSHAP \parencite{lundberg2017unified, lundberg2020treeshap}, we can observe that the estimated parameter $\hat{\mu}$ (denoted ``SEL'' in Figure \ref{fig:simu_feat_imp}) comes as the most important variable of the model.
Other features are as important as their weight from the formula defined in equation (\ref{eq:simu_formula_example}).

Furthermore, we can also see in Figure \ref{fig:simu_pdp} that the model is able to correctly learn the quadratic relation between the SEL variable $\hat{\mu}$ and the target $\mathbf{Y}$.
Our nonlinear model can learn the correct relationship between the SEL variable and the target, which highlights its predominant importance in the model.

% \begin{figure}[!ht]
%    \centering
%    \includegraphics[width=0.8\textwidth]{images/simulation_results.png}
%    \caption{Simulations of SEL with $s = 1$ generated features. The $x$-axis represents the number of features $p$}
%    \label{fig:simu_perf}
% \end{figure}

\subsubsection*{Application to image data}

% https://www.geeksforgeeks.org/visualizing-colors-in-images-using-histogram-in-python/
% http://yann.lecun.com/exdb/mnist/

In order to show the benefit of the SEL methodology on a large spectrum of use cases, we apply it to computer vision data sets.
We use three common data sets for model benchmarking. The first one, the MNIST data set \parencite{lecun-mnisthandwrittendigit-2010}, is composed of $60,000$ images of hand written digits from 0 to 9.
The data set is widely used in machine learning and served as a set for comparing human performance versus machines in reading handwritten digits.
The second one, the Fashion-MNIST \parencite{xiao2017fashion} data set, is a derivative of the original MNIST data with clothes images.
It is also composed of 10 categories of images from T-Shirts, trousers to bags and boots.
The third data set we use is CIFAR-10 \parencite{cifar}.
This database is composed of $60,000$ colored images of animals and vehicles.
It consists of 10 categories ranging from dog, horse to airplane or truck.

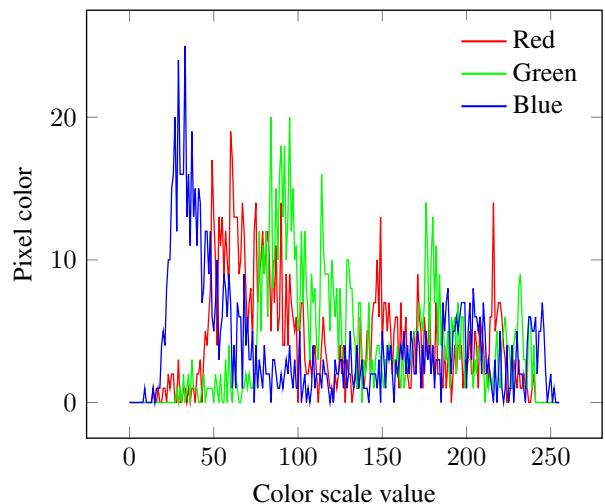
\begin{figure}[!ht]
    \centering
    \begin{tikzpicture}
    \begin{axis}[
    line width=0.5,
    x tick label style={font=\normalsize, draw=none, anchor=north},
    y tick label style={font=\normalsize, draw=none, anchor=east},
    xlabel={Color scale value},
    ylabel={Pixel color},
    label style={font=\normalsize, draw=none},
    legend style={anchor=north east, draw=none},
    %y tick label style={
    %    /pgf/number format/.cd,
    %    fixed,
    %    precision=2
    %},
    ]
        % Distribution for Red
        \addplot[red] table [x=index, y=red, col sep=comma] {images/data/ColorHistogramExample.csv};
        \addlegendentry[draw=none]{Red};
        % Distribution for green
        \addplot[green] table [x=index, y=green, col sep=comma] {images/data/ColorHistogramExample.csv};
        \addlegendentry[draw=none]{Green};
        % Distribution for Blue
        \addplot[blue] table [x=index, y=blue, col sep=comma] {images/data/ColorHistogramExample.csv};
        \addlegendentry[draw=none]{Blue};
    \end{axis}
\end{tikzpicture}
    \caption{Color histogram from horse example image in Figure \ref{fig:cifar_horse}} \label{fig:color_histogram_example}
\end{figure}

\begin{figure}[!ht]
     \centering
     \begin{subfigure}[a]{0.45\textwidth}
         \centering
         \includegraphics[width=0.3\textwidth]{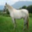}
         \caption{A horse}
         \label{fig:cifar_horse}
     \end{subfigure}
     \hfill
     \begin{subfigure}[a]{0.45\textwidth}
         \centering
         \includegraphics[width=0.3\textwidth]{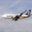}
         \caption{An airplane}
         \label{fig:cifar_plane}
     \end{subfigure}
     \caption{Examples of two different categories in the CIFAR-10 data base}
     \label{fig:cifar_examples}
\end{figure}

To compare our methodology on the aforementioned data bases, we use the same deep learning architecture on the three data sets.
We use the VGG architecture \parencite{simonyan2014very}, which consists of stacking double convolutional layers\footnote{Stack two layers of 2-dimensional convolutions with max-pooling} to process the image before feeding fully connected layers before the final classification.
This Convolutional Neural Network (CNN) architecture  is widely used in computer vision and we will refer to it as the \textit{vanilla} approach. For the SEL approach, we use the same architecture but add more variables on top of the convolutional layers.
For each of our three data sets, we extract the distribution of colors on the images \parencite{novak1992anatomy}.
The MNIST and Fashion-MNIST data being in black and white, the color histogram will correspond to the gray intensity.
The CIFAR-10 being in color, we will extract three histograms from the RGB (red, green and blue) representation of the images.
An illustration of a color histogram for a colored image (displayed in Figure \ref{fig:cifar_horse}) can be found in Figure \ref{fig:color_histogram_example}. From the histogram, we then compute the first four moments (mean, standard deviation, skewness and kurtosis) that will correspond to our SEL 2 features to add to the model.

Our deep learning VGG architecture is only augmented with fully connected layers in parallel of the CNN to ingest the information from the SEL covariates.
If we were to compare the vanilla and SEL models, we can consider the vanilla architecture as a constrained model of our SEL, where weights for the fully connected layers to learning from the moment variables are set to zero.

\begin{table}[!ht]
    \centering
    \begin{tabular}{ccc}
        Dataset & Vanilla & SEL \\
        \midrule
        MNIST & 99.01\% & 99.05\% \\
        Fashion & 90.57\% & 91.17\% \\
        CIFAR-10 & 69.09\% & 69.49\% \\
    \end{tabular}
    \caption{Comparison of model classification accuracy for regular CNN architecture versus SEL augmented model}
    \label{tab:cv_models_perf}
\end{table}

We report the classification performance of both methodologies on our different data sets in Table \ref{tab:cv_models_perf}.
We observe that the SEL features consistently help the model's performance.
Although the accuracy uplift can be modest, such a gain can sometimes be crucial in highly regulated sectors (such as financial institutions, security industry, etc.), where the performance of a model needs to be as high as possible and not acceptable below certain thresholds.
Note that our goal in this exercise is not to reach the least error rate on these data sets (some literature already focuses on this objective by using fine-tuned state-of-the-art model architectures).
Our aim is rather to illustrate that data augmentation via SEL is beneficial to any model, even for a fixed architecture.

We can further observe that the modest uplift of performance on the MNIST data set can be explained by an already high performance of the model, but also by the limited variability of color/gray over the image where the color on the object to detect may be quite monotonous.
As a comparison, the Fashion-MNIST images represent real clothes and objects in their original setting and might have multiple colors.
The gradient of colors can thus be larger, and the SEL features added to the model contain more information. As we can see in Figure \ref{fig:cifar_examples}, the colors in an image of a horse and of an airplane can be very different.
Although one of the goals of the convolutional layers is to recognize the color when analyzing the pixels in the image, having a higher level of information with, for example, moments from the color histogram helps the model eliminate obvious non-candidates more easily.

% 0.0004
% 0.00662
% 0.00579

\subsubsection*{Application to sports data}
%\todo[inline]{Will fill this section soon! Just need to work on the foot data sent by Andreas}

Other applications of data augmented machine learning are in the fields of sports.
\cite{groll2020prediction} compared different regression approaches to model the number of goals scored in international handball games.
They concluded that the Gaussian distribution is the most appropriate distribution to model the number of goals scored by a team and use a LASSO-penalized approach to predict future games.
Taking the same data, we compare the benefit of adding a SEL variable on different models to predict the number of goals scored by the home team during international handball games.

We construct a SEL variable of level 2 using the first moment of the (Gaussian) distribution of past scored goals for each team.
This moment is computed as the average number of goals scored by a team and can be interpreted as a form of strength estimation. We then train a LASSO regression, Random Forest, XGBoost and Neural Network with and without the SEL variable.
The "regular" case corresponds to the same set of features as used in \cite{groll2020prediction}. The case with SEL adds the team's strength with moment estimation on top of these features. We report the RMSE for each model on the training and test set in Table~\ref{tab:hand_model_perf}. As can be seen,  the Random Forest with SEL features performs with the least error on the test set, followed by the LASSO with SEL. We  notice that all models benefit from the adjunction of the SEL covariate.

\begin{table}[!ht]
    \centering
    \begin{tabular}{cccc}
        \toprule
        Model & Feature set & Training & Test \\
        \midrule
        \multirow{2}{*}{LASSO} & Regular & 4.70 & 4.56 \\
        & SEL & 4.05 & 4.39 \\
        \multirow{2}{*}{Random Forest} & Regular & 1.73 & 4.04 \\
        & SEL & 1.69 & 4.00 \\
        \multirow{2}{*}{XGBoost} & Regular & 0.35 & 4.52 \\
        & SEL & 0.33 & 4.47 \\
        \multirow{2}{*}{Neural Net} & Regular & 4.96 & 4.80 \\
        & SEL & 4.89 & 4.73 \\
        \bottomrule
    \end{tabular}
    \caption{RMSE on train and test sets for regular versus SEL models}
    \label{tab:hand_model_perf}
\end{table}

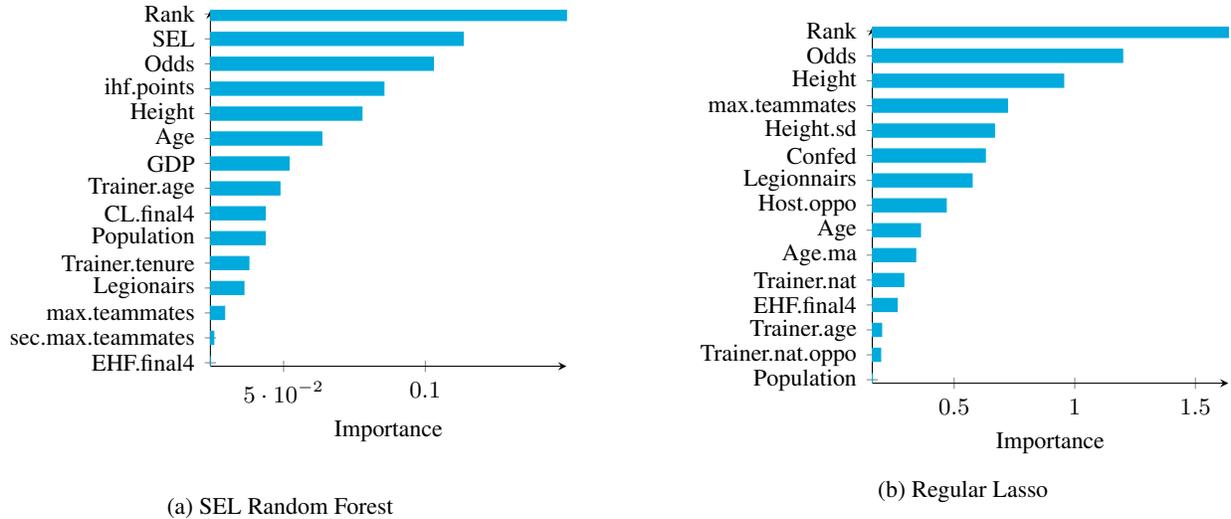
\begin{figure}[!ht]
     \centering
     \begin{subfigure}[a]{0.45\textwidth}
         \centering
         \pgfplotstableread[col sep = comma]{images/data/HandIntlFeatImport.csv}\tabledm
\def\xlistmacro{}
\def\xliststring{}
\pgfplotstableforeachcolumnelement{index}\of\tabledm\as\entry{%
\xifinlist{\entry}{\xlistmacro}{}{
        \listxadd{\xlistmacro}{\entry}
        \edef\xliststring{\xliststring\entry,}
    }
}
\begin{tikzpicture}
    %\begin{axis}[
    % xlabel={Label},
    %ylabel={Count},
    %symbolic x coords/.expand once={\xliststring},
    %xmajorgrids=true,
    %x tick label style={rotate=60,anchor=east}
        %]
        %\addplot [ybar] table [x=feature, y=importance, col sep=comma] {\tabledm};
        %\end{axis}
        \begin{axis}[
            x tick label style={font=\footnotesize, draw=none, anchor=north},
            y tick label style={font=\footnotesize, draw=none, anchor=east},
            label style={font=\footnotesize, draw=none},
            width=0.85\linewidth, % changed size a bit
            height=0.85\linewidth,
            xbar, % not xbar interval
            bar width=5pt,
            %xmin=0.02, % does it make more sense to start at zero?
            %xmax=0.15, % reduced this a bit
            xlabel = Importance,
            axis x line=bottom,
            axis y line=left,
            enlarge y limits=0.01, % no need to enlarge x limits
            symbolic y coords/.expand once={\xliststring},
            ytick = data,
        %    y tick label style={rotate=45, anchor=east}, % I wouldn't use this
            %xticklabel style={
            %  /pgf/number format/fixed % no scientific notation for the smallest values
            %}
         ]
            \addplot[BlueUnilu, fill=BlueUnilu] table [y=index, x=importance, col sep=comma] {\tabledm};
        \end{axis}
\end{tikzpicture}
         \caption{SEL Random Forest}
         \label{fig:feat_import_hand_rf}
     \end{subfigure}
     \hfill
     \begin{subfigure}[a]{0.45\textwidth}
         \centering
         \pgfplotstableread[col sep = comma]{images/data/HandIntlLassoFeatImport.csv}\tabledm
\def\xlistmacro{}
\def\xliststring{}
\pgfplotstableforeachcolumnelement{index}\of\tabledm\as\entry{%
\xifinlist{\entry}{\xlistmacro}{}{
        \listxadd{\xlistmacro}{\entry}
        \edef\xliststring{\xliststring\entry,}
    }
}
\begin{tikzpicture}
    %\begin{axis}[
    % xlabel={Label},
    %ylabel={Count},
    %symbolic x coords/.expand once={\xliststring},
    %xmajorgrids=true,
    %x tick label style={rotate=60,anchor=east}
        %]
        %\addplot [ybar] table [x=feature, y=importance, col sep=comma] {\tabledm};
        %\end{axis}
        \begin{axis}[
            x tick label style={font=\footnotesize, draw=none, anchor=north},
            y tick label style={font=\footnotesize, draw=none, anchor=east},
            label style={font=\footnotesize, draw=none},
            width=0.85\linewidth, % changed size a bit
            height=0.85\linewidth,
            xbar, % not xbar interval
            bar width=5pt,
            %xmin=0.02, % does it make more sense to start at zero?
            %xmax=0.15, % reduced this a bit
            xlabel = Importance,
            axis x line=bottom,
            axis y line=left,
            enlarge y limits=0.01, % no need to enlarge x limits
            symbolic y coords/.expand once={\xliststring},
            ytick = data,
        %    y tick label style={rotate=45, anchor=east}, % I wouldn't use this
            %xticklabel style={
            %  /pgf/number format/fixed % no scientific notation for the smallest values
            %}
         ]
            \addplot[BlueUnilu, fill=BlueUnilu] table [y=index, x=importance, col sep=comma] {\tabledm};
        \end{axis}
\end{tikzpicture}
         \caption{Regular Lasso}
         \label{fig:feat_import_hand_lasso}
     \end{subfigure}
     \caption{Features of importance from SEL Random Forest and Regular Lasso}
     \label{fig:feat_import_hand}
\end{figure}

We further analyze the performance of the Random Forest with SEL features and the regular LASSO by extracting the features of importance in Figure~\ref{fig:feat_import_hand}. We can observe  that the feature \texttt{Rank} is considered by both Random Forest and LASSO models as the most important, and that the new SEL feature comes directly second, showing its high influence.
The features of importance for the Random Forest have been extracted using the built-in module in Scikit-learn, while for the LASSO we have derived the importance by taking the absolute value of the estimated coefficients presented in Table 4 from \cite{groll2020prediction}.

The order of important variables between the original model of \cite{groll2020prediction} and our fitted Random Forest is respected; adding a SEL variable of level 2 helps reduce the error of the model while the feature is considered as highly important.

\section*{Code availability}\label{sec:code}

The code and software materials have been deposited to the GitHub page at \url{https://github.com/florianfelice/StatisticallyEnhancedLearning}.

\printbibliography

\end{document}